\documentclass[aps,preprintnumbers,superscriptaddress,twocolumn,amsmath,amssymb,floatfix,prb]{revtex4} 
\pdfoutput=1
\usepackage{graphicx}
\usepackage[usenames,dvipsnames]{color}
\usepackage[colorlinks,bookmarks=false,citecolor=NavyBlue,linkcolor=Red,urlcolor=blue]{hyperref}
\usepackage{float}

\usepackage[export]{adjustbox}

\usepackage{lmodern}
\usepackage{graphicx}% Include figure files

\usepackage[export]{adjustbox}
\usepackage{dcolumn}% Align table columns on decimal point
\usepackage{bm}% bold math
\usepackage{hyperref}% add hypertext capabilities
\hypersetup{linktocpage,colorlinks,citecolor={blue},pdfdisplaydoctitle=true,pdfpagemode=UseOutlines,bookmarksnumbered=true}
\usepackage{mathrsfs}
\usepackage{xcolor}
\usepackage[normalem]{ulem}

\usepackage{comment}

\newcommand{\haar}{\mbox{CUE}}
\newcommand{\round}[1]{\lfloor #1 \rceil}

\newcommand{\dt}{\tau}
\newcommand{\bra}[1]{\langle #1 |} 
\newcommand{\ket}[1]{| #1 \rangle }

\definecolor{cbl}{rgb}{0,0,1}
 
\definecolor{crd}{rgb}{1,0,0}

\newcommand{\dd}{\text{d}}

\newcommand\be{\begin{equation}}
\newcommand\bea{\begin{eqnarray}}
\newcommand\bes{\begin{subequations}}
\newcommand\esu{\end{subequations}}
\newcommand\ee{\end{equation}}
\newcommand\eea{\end{eqnarray}}

\newcommand{\cmmnt}[1]{}

\newcommand{\vqp}{v_{QP}}
\newcommand{\vlr}{v_{LR}}
\newcommand{\vB}{v_{\rm B}}

\newcommand{\vmax}{v_{\rm max}}

\newcommand{\ltqp}{\tau}

\newcommand{\svn}{\mathcal{S}}

\def\doi{http://dx.doi.org/}

\usepackage{braket}

\newcommand\ocite[1]{[\onlinecite{#1}]}

\newcommand{\titleinfo}{Entanglement front generated by an impurity 
travelling in an isolated many-body quantum system}

\begin{document}

\title{\titleinfo}

\author{Andrea De Luca}
\affiliation{Laboratoire de Physique Th\'eorique et Mod\'elisation (CNRS UMR 8089), Universit\'e de Cergy-Pontoise, F-95302 Cergy-Pontoise, France}
\affiliation{The Rudolf Peierls Centre for Theoretical Physics, Oxford University, Oxford, OX1 3NP, United Kingdom}

\author{Alvise Bastianello}
\affiliation{Institute for Theoretical Physics, University of Amsterdam, Science Park 904, 1098 XH Amsterdam, The Netherlands}

%\date{\today}

\begin{abstract}
We investigate the effect on the entanglement dynamics of an impurity moving at constant velocity in a closed quantum system. We focus on one-dimensional strongly-correlated lattice models, both in the presence of integrable and chaotic dynamics. In the former, the slow impurity is preceded by fast quasiparticles carrying an ``\textit{endogenous}'' entanglement front which decays in time as a power-law; on the contrary, a fast impurity drags itself an ``\textit{exogenous}'' entanglement front which never fades.
We argue that these effects are valid for generic systems whose correlations propagate inside a light-cone. To assess the fully chaotic regime, we formulate a random circuit model which supports a moving impurity and a sharp lightcone. Although the qualitative behavior is similar to the integrable case, the endogenous regime is only visible at short times due to the onset of diffusive energy transport.
Our predictions are supported by numerical simulations in the different regimes.
\end{abstract}

\maketitle
%\tableofcontents

\section{Introduction}

Understanding the out-of-equilibrium dynamics of isolated many-body quantum systems is one of the main challenges of current research in low-energy physics. A fundamental question is how the statistical description of thermodynamics could emerge via coherent quantum evolution \ocite{cit_eq} and a crucial ingredient is provided by entanglement \ocite{cit_endef}, which reflects the amount of quantum correlations between different portions of the system. Generically, weakly entangled initial states show a fast growth of quantum correlations: the initial information is rapidly encoded in non-local degrees of freedom and thermodynamics emerges. Entanglement is generated by the spreading of correlations, which, in many systems, happens with a finite maximal velocity, e.g. because of the Lieb-Robinson bound~\ocite{liebrob}. 

In a pure state, a basic way to quantify the amount of quantum correlation of a region $\mathcal{A}$ with the rest $\bar{\mathcal{A}}$  is to employ entanglement entropy $S_{\mathcal{A}}$, defined from the reduced density matrix as $S_\mathcal{A}=-\text{Tr}\rho_\mathcal{A}\log\rho_\mathcal{A}$. In 1D, a particularly effective and simple picture to understand its behaviour is provided by the quasiparticle interpretation.
 First introduced in conformal field theories~\ocite{cc-05,CC:review,eh-cft,WeRy18} and then confirmed in free~\ocite{fagottiXY,ep-08,nr-14,bkc-14,coser-2014,buyskikh-2016,cotler-2016,betaca18,BaCa18} and integrable models~\ocite{alba2017,alba-2018,p-18}, it has provided valuable insights even beyond these settings~\ocite{confinement2017}. 
In this picture, when a portion of a system is brought out-of-equilibrium, pairwise entangled quasiparticles are produced at each point in space (see however \ocite{betaca18,BaCa18}), which propagate  through the system in opposite directions:  as soon as two entangled quasiparticles get to be shared between $\mathcal{A}$ and  $\bar{\mathcal{A}}$, the amount of entanglement between the two subsystems is increased.
For many systems, Lieb-Robinson provides an upper bound $\vlr$ for the maximal quasiparticle velocity $\vqp < \vlr$ which induces an effective \textit{horizon}. As a consequence, the entanglement entropy $S_{\mathcal{A}}$ grows linearly in time, up to saturation to a value proportional to $\ell_{\cal A}$, the length of $\cal A$.

The existence of an horizon has been observed in experiments~\ocite{cheneau2012, jurcevic2014, schmiedmayer2013} and in models where no quasiparticles can be consistently defined~\ocite{ballistic2013}. Recently, solvable models of chaotic dynamics based on random circuits have established the existence of well-defined lightcones. They also provided a quantitative characterisations of different velocities, as they emerge from entanglement growth and operator spreading~\ocite{kpznahum,operatornahum, rucu1, rucdiffusion}. 

\begin{figure*}[t!]
\includegraphics[width=1\textwidth]{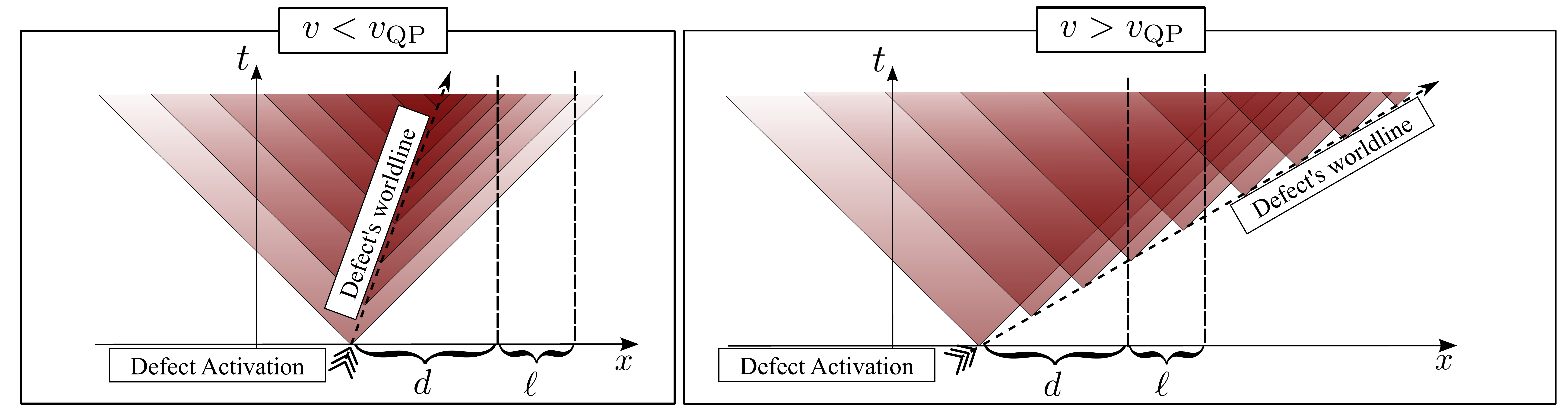}
\caption{\label{fig_envelope}\emph{Quasiparticle envelop for, respectively, a subluminal and superluminal defect. 
The dashed arrow represents the moving defect, which is activated in correspondence with the tail of the arrow and then moves in a straight line in the space-time plane. During its motion, it emits burst of excitations (red cones) that then freely propagate. The lengths $\ell$ and $d$ are depicted for convenience of the forthcoming entanglement's growth discussion (see Fig. \ref{fig_EE_growth_free}-\ref{fig_entanglement_int},\ref{entRUC}).
As it is clear from the picture, the nature of the propagating front is remarkably different in the two cases: while in the subluminal defect the edge is increasingly damped 
as it propagates, this is no longer true for the superluminal case where the wavefront is continuously supplied with new excitations.
}}
\end{figure*}

A very natural test of these velocities consists in injecting in the system a localized perturbation which travels at constant velocity $v$ \ocite{AA2018,cit_movingimp,AA2018ising,BuCh014,GaLy14,Ga14}. If $\vmax \lesssim \vlr$ defines the effictive maximal velocity at which signals can propagate in the quantum system under examination, a dramatic difference between the $v\ll \vmax$ and $v\gg\vmax$ regimes is expected, which we refer to as \textit{subluminal} and \textit{superluminal} case, respectively. 
Relativistic analogies of this kind have already been realized in condensed matter setups, mostly as an emergent description of low-energy excitations \ocite{unruh,carusotto1,carusotto2,sondhi18}. While traveling localized perturbations have been considered in several contexts \ocite{refB1,refB2,refB3,refB4,AsPi04,BuGe01}, the key role played by the maximal velocity has been unveiled only recently \ocite{WoCa10,sondhi18,AA2018,AA2018ising,cit_movingimp,BuCh014,GaLy14,Ga14}.
In this paper, we investigate in generality the implications drawn from the presence of a finite $\vmax$, exploring the hallmarks of integrability and chaotic dynamics.

We focus on one dimensional, short-range lattice systems where $\vlr$ is finite and efficient numerical simulations are possible through \textit{matrix-product states} techniques~\ocite{MPSrev,karrasch2015}. We first consider a spin chain which can be mapped onto non-interacting fermions by the use of a Jordan-Wigner transformation, then consider a travelling impurity and show that the entanglement phenomenology exhibits dramatic differences whether the velocity of the impurity $v$ is above or below the one of quasiparticles. In particular, we show that a fast impurity carries a front of entanglement which never fades. 
Our numerical simulations show that this phenomenology remains robust at the numerically accessible timescales, even in the presence of weak integrability breaking perturbations which result in a finite quasiparticle lifetime $\ltqp$. On larger time scales $t \gg \ltqp$, diffusive transport is expected to dominate: the system response is always slower than the impurity for arbitrary $v$. 
This is confirmed by a random unitary circuit that we introduce, where a $U(1)$ charge is conserved everywhere except at the position of the moving defect. 

The paper is organized as follows: in Sec. \ref{sec_quasiparticles} we consider quantum spin chains, both integrable and with weak integrability-breaking terms. In this regime, the quasiparticle picture holds and the features of the entanglement spreading can be understood within this framework, giving quantitative predictions for the free case. 
Sec. \ref{sec_random} is instead dedicated to the study of a random unitary circuit supporting a $U(1)$ conservation law, locally broken by the moving defect. In this case, no quasiparticles can be consistently defined, but a similar phenomenology emerges at intermediate times. At long times, transport is dominated by diffusion which suppresses entanglement propagation. 
The presence of a persistent entanglement front in the superluminal case regardless the applicability of the quasi-particle picture indicates a universal mechanism. Indeed, in Sec. \ref{sec_comovingtail} we show how a superluminal defect immediately generates a comoving steady state, rigidly following the impurity, whose size grows linearly in time. Our argument is based solely on the existence of a maximum velocity, thus it is of widest applicability.
The generation of such a comoving steady state explains the persistence of the entanglement front and is clearly visible in the profiles of local observables.
Our conclusions are then gathered in Sec. \ref{sec_conclusions}. Numerical methods are presented in the appendix.

\newpage
\section{Moving defect and quasiparticles}
\label{sec_quasiparticles}

We consider the paradigmatic example of the Ising spin chain
\be
H=-\frac{1}{2}\sum_{j=1}^N \sigma_j^x\sigma_{j+1}^x+(h_z + V(j - vt))\sigma_j^z + h_x \sigma_j^x \, ,\label{Hising}
\ee
which encompasses a broad phenomenology including integrable ($h_x = 0$) and non-integrable $(h_x\ne 0)$ dynamics, with a second-order phase transition ($h_z = 1$, $h_x = 0$). The potential $V(x)$ describes a perturbation in the transverse magnetic field, which travels at constant velocity $v$. We focus on the simplest case of an extremely narrow perturbation, i.e. $V(x) = \kappa \, \delta(x)$, but our conclusions hold in the general setting of localized potentials $V(x)$. The Ising spin chain has also been realized in cold-atom experiments~\ocite{isingcold} and moving perturbations as in Eq. \eqref{Hising} could be realized as a travelling impurity~\ocite{tonksimpurity}, or as moving spin flip \ocite{fukuhara2013}. In particular, the choice of the $\delta-$defect best describes this latter possibility.

Let us first discuss the integrable point. At $h_x = 0$, the protocol can be exactly solved \ocite{cit_isingrew, AA2018ising}. Within the quasiparticle picture, the moving impurity can be regarded as emitting bursts of excitations during its motion (Fig. \ref{fig_envelope}) \ocite{AA2018, AA2018ising} (see also Ref. \ocite{DeLu14,BeFa16,Fa15}). 
More specifically, in the absence of the external potential ($\kappa=0$), the Hamiltonian $H$ \eqref{Hising} can be diagonalized in Fourier space, combining a Jordan-Wigner (JW) transformation and a Bogoliubov rotation. 
The JW introduces fermionic degrees of freedom with standard anticommutation rules $\{d_j,d_{j'}\}=\delta_{j,j'}$, where
\be
d_j=e^{i\pi\sum_{l=1}^{j-1}\sigma^+_l\sigma_l^-}\sigma^+_j\, ,\label{JW}
\ee
and $\sigma_l^\pm=(\sigma^x_l\pm i\sigma^y_l)/2$. At $h_x=0$, the Hamiltonian \eqref{Hising} in the new basis is readily expressed as
\be
H=\sum_{j=1}^N-\frac{1}{2}\left(d^\dagger_j d^\dagger_{j+1}+d^\dagger_jd_{j+1}+\text{h.c.} \right)+[h_z+V(j-vt)]d^\dagger_jd_j\, .\label{Hfermion}
\ee
Above, ``h.c." stands for the hermitian conjugated of the expression in brackets. If $h_x\ne 0$, the fermions become interacting and the model is no longer exactly solvable.
At finite size, periodic boundary conditions on the spin chain induce (anti)periodic boundary conditions in the (even)odd magnetization sectors in the fermionic basis. 
However, we are ultimately interested in the thermodynamic limit and this complication can be safely neglected.
In the absence of the defect ($V=0$), the Hamiltonian \eqref{Hfermion} is readily diagonalized in the Fourier space via a Bogoliubov rotation 
\be\label{eq:mode}
\begin{pmatrix} d_j \\ d^\dagger_{j}\end{pmatrix}=\int_{-\pi}^\pi \frac{{\rm d}p}{\sqrt{2\pi}}\, e^{ipj}\begin{pmatrix}\cos\theta_p && i\sin\theta_p \\ i\sin\theta_p && \cos\theta_p  \end{pmatrix}\begin{pmatrix}\alpha_p \\ \alpha^\dagger_{-p}\end{pmatrix}\, ,
\ee
The fermionic operators $\alpha_p$ satisfy canonical anticommutation rules $\{\alpha^\dagger_p,\alpha_q\}=\delta(p-q)$  and diagonalize the Hamiltonian as
\be
H=\int_{-\pi}^\pi \frac{{\rm d}p}{2\pi}\, \omega(p)\alpha^\dagger_p\alpha_p+\text{const.}
\ee
where $\omega(p)=\sqrt{(\cos p-h_z)^2+\sin^2p}$, provided the Bogoliubov angle $\theta_p$ is chosen as
\be
\tan\theta_p=\frac{\omega(p)+\cos p-h_z}{\sin p}\, .
\ee
The modes $\alpha_p$ are readily interpreted as the entangling quasiparticles that are thus moving with velocity $v(p)=\partial_p \omega(p)$. In the free case, the maximum velocity of the free modes set the maximum velocity of the spreading of quasiparticles and operators, i.e.
\be
v_{QP}=\max|v(p)|=\min(|h_z|,1)\, .
\ee

For definiteness, we consider the system initially prepared at $\kappa = 0$ in the paramagnetic groundstate ($h_z > 1$), which is only weakly entangled because of area law \ocite{cit_arealaw}.
At $t>0$, the moving perturbation $\kappa > 0$ is activated and excitations are created on top of the initial ground state.
Physically, due to the locality of the perturbations, excitations are locally emitted  from the moving defect and then freely propagate in the bulk: this is indeed confirmed by an exact solution of the protocol.
Despite the model \eqref{Hfermion} being free, its exact solution in the presence of the defect is not trivial and requires rather lengthy calculations, which have been presented in Ref. \ocite{AA2018ising} (see also Ref. \ocite{AA2018}). Here, we just quote the results we need for our purposes, leaving to the original reference their derivation.
In the scaling region far from the defect, correlation functions of local observables can be derived in terms of a space/time inhomogeneous mode density. For example, the local fermionic density is
\be\label{eq_local_dens}
\langle d^\dagger_j(t) d_j(t)\rangle=\int_{-\pi}^\pi\frac{\dd p}{2\pi}(\sin^2\theta_p+\cos(2\theta_p)\eta_{j,t}(p))\, .
\ee
At fixed time and position, this is the expectation value we would have derived in the Homogeneous Ising Hamiltonian on a state diagonal on the modes $\alpha_p$ and with mode density $\langle \alpha^\dagger_p\alpha_q\rangle=\delta(p-q)\eta_{j,t}(p)$. Eq. \eqref{eq_local_dens} can be extended to arbitrary expectation values of local observables.
Semiclassically, we can interpret $\eta_{j,t}(p)$ as a local phase-space density of the excitations generated by the moving defect. Indeed, its space-time evolution can be recast as
 \ocite{AA2018ising}
\be\label{etapos}
\eta_{j,t}(p)=\Theta[(j-vt) (v(p)-v)] \Theta[|(v(p)-v)t|-|j-vt|]\eta_\text{scat}(p)\, ,
\ee
where $\Theta$ is the Heaviside Theta function and $\eta_\text{scat}(p)$ is the density of quasiparticles produced by the impurity.
The function $\eta_\text{scat}(p)$ encodes all the dynamics and depends on the specific shape of the defect. 
The Theta functions in Eq. \eqref{etapos} convey a simple physical message: excitations at a given $p$ are present only ``beyond" the defect (where ``beyond" is decided by the sign of the relative velocity through the Theta function $\Theta[(j-vt) (v(p)-v)]$) and up to the maximal distance that the flux of particles can have reached.
It must be stressed out the peculiar form of Eq. \eqref{etapos}, indeed $\eta_{j,t}$ depends on time and position only through the combination $\zeta=j/t$ and is therefore scale-invariant.

For the $\delta-$defect considered here, an exact analytical computation of $\eta_\text{scat}(p)$ is possible (see \ocite{AA2018ising} for details). Here, we report the resulting expression in the simpler case of the superluminal defect
\begin{multline}\label{delta_scat}
\eta_\text{scat}(p)=\left|\frac{{\rm v}_1(p)}{{\rm v}_2(p_2)}\right|\Big|u^\dagger_1(p)\Big[ e^{i\frac{\kappa}{v}\sigma^z}\mathcal{K}^-(p_2)-\mathcal{K}^+(p_2)\Big]^{-1}\times\\ (1-e^{i\frac{\kappa}{v}\sigma^z})u_{2}(p_2)\Big|^2\, ,
\end{multline}
where we denote
\be\label{eq_udef}
u_1(p)=\begin{pmatrix}\cos\theta_p  \\ i\sin\theta_p   \end{pmatrix}\,,\hspace{2pc}u_2(p)=\begin{pmatrix} i\sin\theta_p \\ \cos\theta_p  \end{pmatrix}\, ,
\ee
and ${\rm v}_1(p)=v(p)-v$ and ${\rm v}_2(p)=-v(p)-v$. The value 
$p_2$ is instead defined as the (unique) solution $p_2\ne p$ of $\epsilon_1(p)=\epsilon_2(p_2)$ where
\be
\epsilon_{1}(p)=\omega(p)-vp\,\hspace{3pc} \epsilon_{2}(p)=-\omega(p)-vp\, .
\ee
Finally, $\mathcal{K}^\pm$ are $2\times2$ matrices defined as it follows
\begin{multline}
\mathcal{K}^\pm(p)=\pm\frac{i}{2v}+\sum_{b=1,2}\frac{u_b(p_{b})\,u^\dagger_b(p_{b})}{2i|{\rm v}_b(p_{b})|}\\
+\mathcal{P}\int_{-\infty}^\infty \frac{{\rm d}q}{2\pi} \sum_{b=1,2}\frac{u_b(q)\,u^\dagger_b(q)}{\epsilon_2(p)-\epsilon_b(q)}\, ,
\end{multline}
where we set conventionally $p_1=p$ and use the Principal Value prescription to handle the singularity in the integral. A similar, albeit more complicate expression, is available also in the subluminal case \ocite{AA2018ising}.
If $v<\vqp$,  a fraction of the emitted quasiparticles is faster than the perturbation (see Fig. \ref{fig_envelope}). 

The wavefront results from the fastest excitations $v\sim \vqp$: there will be a finite number of particles emitted with a velocity in $[\vqp - \Delta v, \vqp]$, which, after a time $t$, will be spread on a growing length $\Delta x = \Delta v \, t$.
As a consequence, the propagating front of the perturbation becomes weaker and weaker as time grows, with a power law decay.

\begin{figure}[t!]
\begin{center}
\includegraphics[width=1\columnwidth,valign=r]{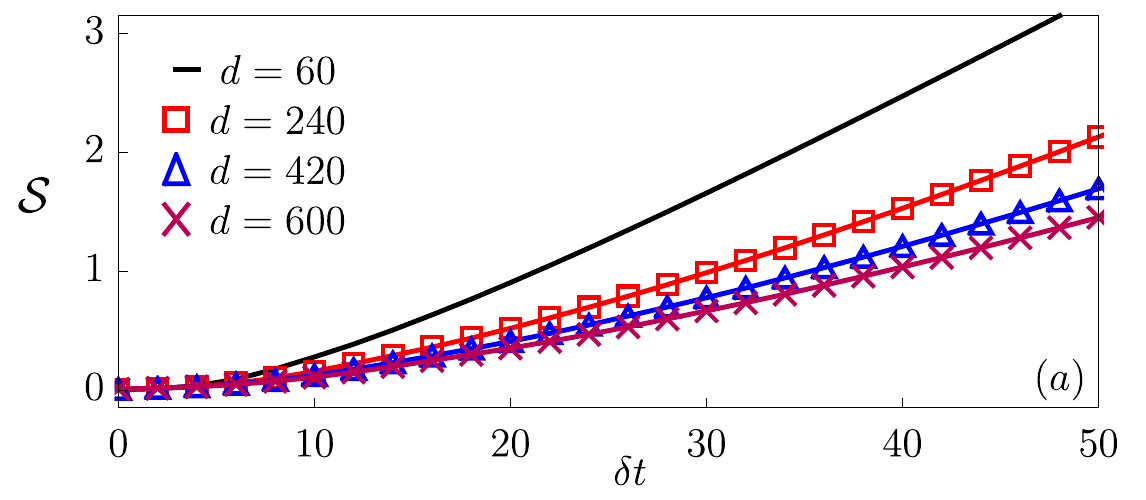}
\includegraphics[width=1\columnwidth,valign=r]{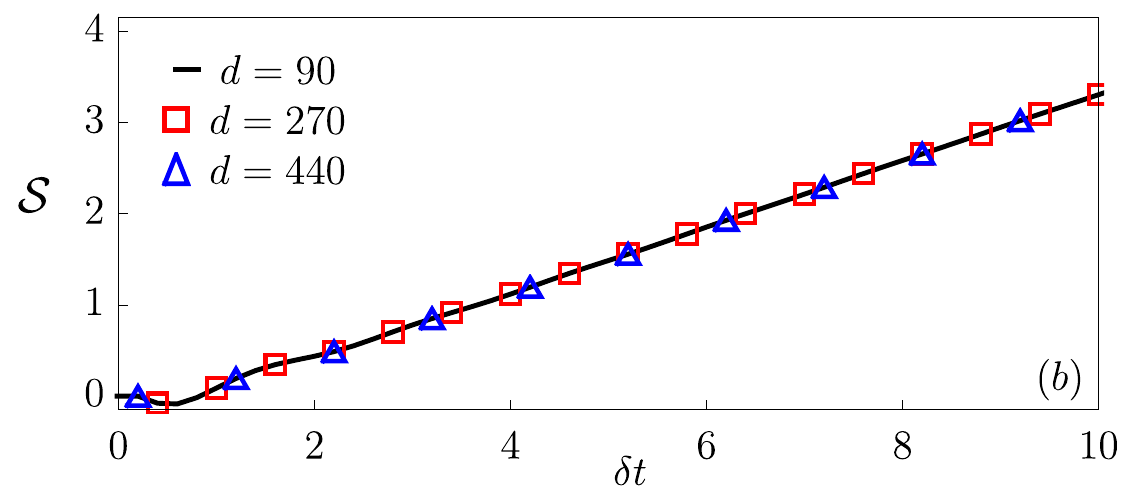}
\caption{\label{fig_EE_growth_free}
\emph{
The entanglement entropy of the spacial region $[d, \infty)$ is plotted for several values of $d$ against the time $\delta t = t - \min(d/\vqp, d/v)$, elapsed after the wavefront enters the region.
Panel $(a)$: subluminal defect ($v=0.5$, $h_z=1.1$) at the integrable point $h_x=0$.  Panel $(b)$: superluminal defect ($v=5$, $h_z=1.1$) at integrable point $h_x=0$. Compared with $(a)$, no damping is observed.
}}
\end{center}
\end{figure}

This feature is clearly displayed in the growth of the Entanglement Entropy, see Fig. \ref{fig_EE_growth_free}.
We performed numerical simulations employing a Green function method which allowed us to efficiently simulate the dynamics induced by Eq. \eqref{Hising} for large systems and long times (see App. \ref{app_numerics}). 
In Fig.~\ref{fig_EE_growth_free} we consider half lines $[d, \infty)$ placed on the right of the perturbation, at increasingly larger distances $d$ from its initial position (see also Fig.~\ref{fig_envelope}). In the subluminal case Fig. \ref{fig_EE_growth_free} (panel $(a)$) The entanglement growth is clearly slower and slower as the distance is increased.

The picture in the superluminal case ($v>\vqp$) is completely different (see  Fig.~\ref{fig_EE_growth_free} (panel $(b)$), since
the defect continuously generates new quasiparticle excitations, whose wavefronts stockpile behind the defect itself (see Fig.~\ref{fig_envelope} right).
Indeed, this difference is clearly reflected on the growth of entanglement entropy: the rate at which entanglement grows does not experience any damping if the distance of the halfline is increased.

\begin{figure}[b!]
\begin{center}
\includegraphics[width=1\columnwidth,valign=r]{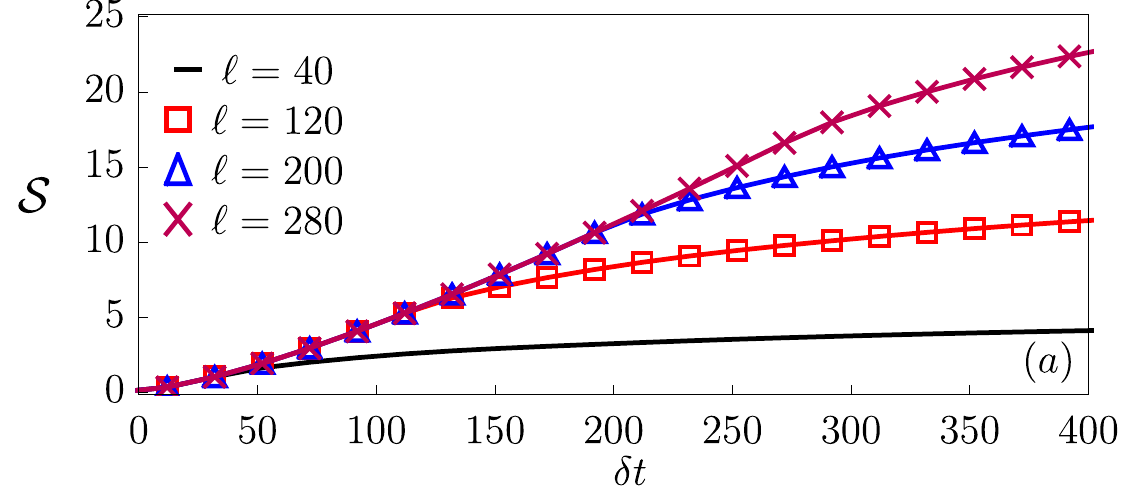}
\includegraphics[width=1\columnwidth,valign=r]{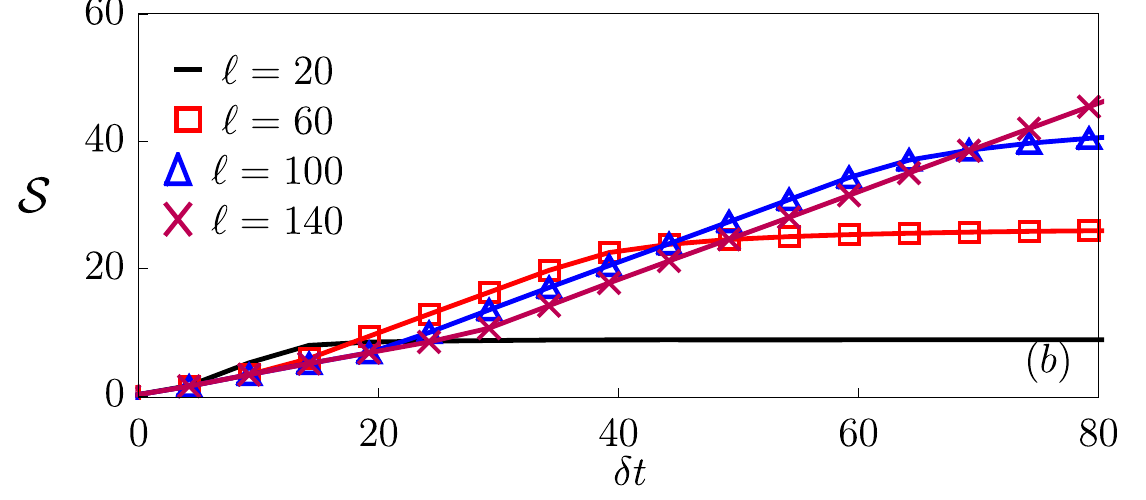}
\includegraphics[width=1\columnwidth,valign=r]{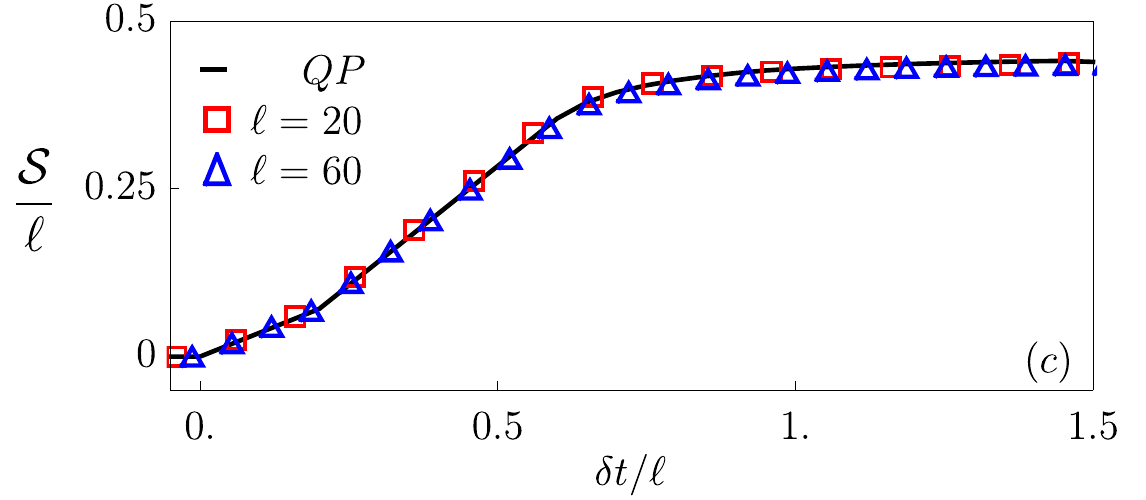}
\caption{\label{fig_EE_growth_free_int}
\emph{
Entanglement growth in the Ising chain at the integrable point $h_x=0$ and saturation for different intervals of size $\ell$, all posed at the same distance $d$ from the defect, being the latter subluminal (Subfigure $(a)$, $h_z=1.1$ $v=0.5$) and superluminal (Subfigure $(b)$ $h_z=1.1$ $v=5$). In the subluminal case the initial growth is not linear and diminishes its growth rate after the wavefront has covered the whole interval, i.e. after a time $\delta t\sim \ell/\vqp$. 
In the superluminal case, the frontwave overcomes the interval at $\delta t=\ell/v$ and the slope experiences a sudden increment, which is well understandable within the quasiparticle picture and found to be an exact doubling of the growth rate. 
Panel $(c)$: comparison of the scaling form of the entanglement entropy against the quasiparticle picture Eq. \eqref{EE_analytic} in the superluminal case (Subfigure $(b)$ $h_z=1.1$ $v=5$). For times $t\ll d/\vqp$, the entanglement growth is a scaling form in terms of the size of the interval $\mathcal{S}=\ell f(\delta t/\ell)$, which is depicted.
}}
\end{center}
\end{figure}

A further difference between the superluminal and subluminal case can be observed looking at the entanglement entropy of a finite interval $[d, d+\ell]$, placed far away on the right of the defect (Fig.~\ref{fig_EE_growth_free_int}).
The interval is first hit by the wavefront which, at larger times, completely overcomes the interval (Fig.~\ref{fig_envelope}).
In the subluminal case, we assume $d \gg \ell \gg 1$, so that the defect only reaches the interval at much later times and we can focus only on the effect of the propagating wavefront. 
As the wavefront proceeds through the interval, the entanglement grows as in the half-line case previously considered: no signal made its way to the right endpoint of the interval
which therefore does not play any role.
The main difference appears when the wavefront leaves the interval: indeed, in the subluminal case 
the entanglement growth rate diminishes (Fig.~\ref{fig_EE_growth_free_int} (panel $(a)$). This is due to the progressive saturation of the entanglement carried by the quasiparticles at different velocities: only slower and slower quasiparticles keep contributing to the generation of entanglement. 

The superluminal case displays the opposite trend (Fig. \ref{fig_EE_growth_free_int} (panel $(b)$): the entanglement entropy enhances its growth with a sudden change in the slope, progressively diminishing again its growth only at a later stage. 
In fact, the superluminal defect continuously generates entangled quasiparticles and when it overcomes the interval, the excitations start to entangle the interval also across the right edge (see Fig.~\ref{fig_envelope}). This contribution equals the one coming from the left edge, exhibiting therefore a precise doubling of the entanglement growth rate. 

\subsection*{Analytic calculation of entanglement entropy}
All these features can be quantitatively captured through a proper generalization of the quasiparticle picture.
In its original formulation, the quasiparticle picture applies to homogeneous quenches (for recent generalizations to inhomogeneous setups see Ref. \ocite{BeFaPiCa18, AlBeFa19,Alba18}) with a pair structure of the initial state in the post-quench basis (see however Ref. \ocite{betaca18, BaCa18} for generalizations beyond the pair structure): after the quench, excitations are locally produced in pairs of opposite momentum which subsequently travel ballistically across the system. The entanglement between the regions $A$ and $\bar{A}$ receives contributions only from those pairs that are shared among the two parts, namely at time $t$ one quasiparticle of a pair belongs to $A$ while the other to $\bar{A}$. This configuration contributes with some weight $s(p)$ dependent on the momentum. The total entanglement is just the sum of the contributions coming from each pair.
As long as single pairs of excitations in free systems are concerned, the contribution to the entanglement $s(p)$ can be found by a consistency requirement with the emergent stationary state \ocite{fagottiXY} and is completely fixed by the excitation density $\eta(p)$
\be\label{eq_s_weight}
s(p)=-\eta(p)\log\eta(p)-\left[1-\eta(p)\right]\log\left[1-\eta(p)\right]\, .
\ee
We now see how this picture can be promptly generalized to describe the protocol involving a moving defect considered here.
In this respect, it is important to note that, since the Hamiltonian is clearly quadratic in the fermionic basis, quasiparticles must be created and destroyed pairwise. However, while in a homogeneous quench (i.e. translational invariant) quasiparticles within the same pair are produced with opposite momenta, in the moving defect framework this does not hold true any longer.
Consider a change of reference frame and set the defect at rest. In this viewpoint, the initial state is moving with velocity $-v$ and the particle creation can be most easily understood within a scattering framework: the initial vacuum flows towards the defect and scatters, producing then pairwise excitations.
In the comoving reference frame, the Hamiltonian looses any explicit time dependence: the energy is conserved and thus the pair of particles must have the same total energy as the vacuum, which is of course zero. In this argument, we must use the energy in the comoving reference frame $\epsilon_1(p)=\omega(p)-vp$.
Thus, if a pair of excitations with momenta $(p,\bar{p})$ is produced it must hold true
\be\label{enconservation}
\epsilon_1(p)+\epsilon_1(\bar{p})=0\, .
\ee
Notice that in general $p\ne -\bar{p}$, though this is recovered in the limit of an infinitely fast defect $v\to \infty$. This is expected, since an extremely fast defect excites simultaneously the whole system and is therefore equivalent to a sudden global quench.

If the impurity is superluminal, the energy conservation \eqref{enconservation} possesses a unique solution \ocite{AA2018ising} and the well-established quasiparticle picture \ocite{fagottiXY} can be straightforwardly generalized. On the contrary, in the case where the defect is subluminal Eq. \eqref{enconservation} possesses more than a solution: particles are still produced pairwise, but the outgoing state is a quantum superposition of all different pairs compatible with energy conservation.
Hereafter, we focus solely on the superluminal defect case, but the forthcoming computation of the entanglement growth can be generalized to the subluminal case along the line of Ref. \ocite{BaCa18}.
The defect is a source of quasiparticles emitted in pairs $(p,\bar{p})$: quasiparticles belonging to the same pair are entangled with each other, while quasiparticles of distinct pairs are disentangled, as well as particles emitted at different positions and times.
After being produced, the particles within a pair travel with constant velocities $v(p)$ and $v(\bar{p})$ respectively, carrying entanglement through the system.

In particular, consider an interval of extrema $\mathcal{A}=[d,d+\ell]$: the entanglement between the interval $\mathcal{A}$ and its complementary $\bar{\mathcal{A}}$  will receive contributions only from those pairs of quasiparticles such that, within the same pair, one quasiparticle lays in $\mathcal{A}$ and the other in $\bar{\mathcal{A}}$.
The computation of the entanglement entropy ultimately boils down to the very geometric problem of counting how many pairs will contribute, associating to each pair $(p,\bar{p})$ the proper weight $s_\text{scat}(p)=s_\text{scat}(\bar{p})$.

\begin{multline}\label{EE_analytic}
\mathcal{S}(t)=\int_{-\pi}^\pi \frac{{\rm d} p}{2\pi}\,\int_{-\infty}^{vt} {\rm d } y\,\, \chi[y+(t-y/v)v(p)]\times\\
\bar{\chi}[y+(t-y/v)v(\bar{p})]\,s_\text{scat}(p)\, .
\end{multline}
Above, $\chi$ is the characteristic function of the interval $\mathcal{A}$, i.e.
\be
\chi(j')=\begin{cases} 1\hspace{2pc} j'\in \mathcal{A}\\
0\hspace{2pc} j'\notin \mathcal{A}\end{cases}\, ,
\ee
while the function $\bar{\chi}(x)=1-\chi(x)$ is simply the one of $\bar{\mathcal{A}}$. The above formula is easily interpreted: pairs of quasiparticles are originated in position $y$ because of the passage of the defect and this happens at a time $y/v$. Subsequently, each of the two particles freely travel reaching at time $t$, respectively, position $y+(t-y/v)v(p)$ and $y+(t-y/v)v(\bar{p})$. The pair will contribute to the entanglement if one of the particles lays within $\mathcal{A}$ and the other in $\bar{\mathcal{A}}$: this is ensured by the product of characteristic functions.
The correct weight $s_\text{scat}$ can be fixed from the homogeneous result Eq. \eqref{eq_s_weight} replacing the excitation density with that produced by the defect
\eqref{etapos}, i.e. $\eta(p)\to \eta_\text{scat}(p)$. The fact that $\eta_\text{scat}(p)=\eta_\text{scat}(\bar{p})$ ensures that $s(p)=s(\bar{p})$, as it should be.
In Fig. \ref{fig_EE_growth_free_int} (panel $(c)$) we provide the comparison between the numerical data and the analytical prediction of the quasiparticle picture, finding perfect agreement: for times $t\ll d/\vqp$, the quasiparticle predicts for the entanglement growth a scaling function $\mathcal{S}(t)=\ell f(\delta t/\ell)$, with $\delta t$ the time lapse with respect to the moment when the interval is first hit by the wavefront.

\subsection*{Effect of integrability breaking}
We now leave the integrable point and see how the previous picture is affected by a small integrability breaking term $h_x\ne 0$: in this case, quasiparticles are no longer stable and must acquire a finite lifetime $\tau$. At $t \gg \tau$, they undergo a complex dynamics which eventually leads to thermalization~\ocite{bertini2015}. However, note that this large-time regime is not accessible by current numerical simulations based on \textit{tensor network methods} (App. \ref{app_numerics}), as entanglement generated by the defect becomes too large. 
Despite the lack of analytical prediction and the finite life-time,  quasiparticles still provide a great insight about the entanglement production.

\begin{figure}[t!]
\begin{center}
\includegraphics[width=1\columnwidth,valign=r]{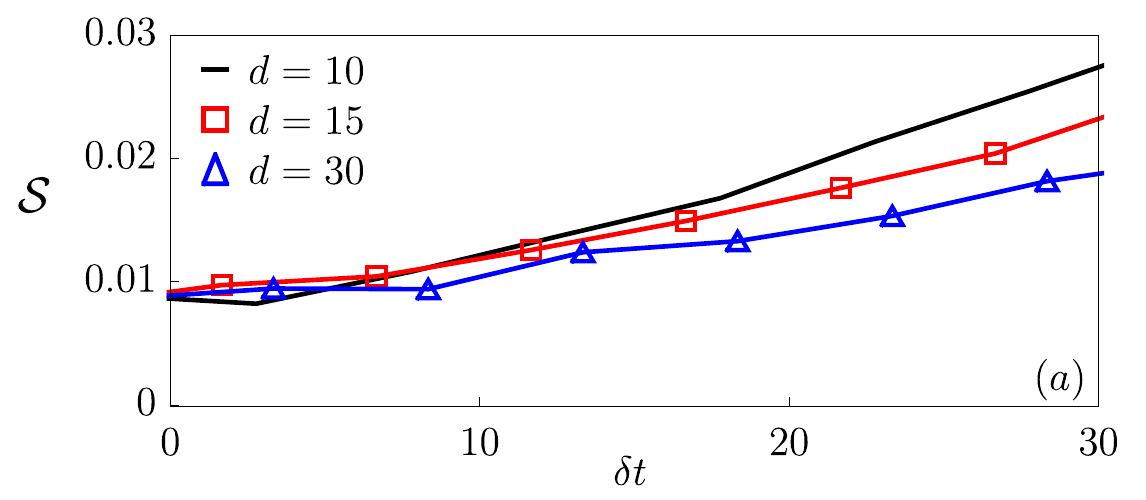}
\includegraphics[width=1\columnwidth,valign=r]{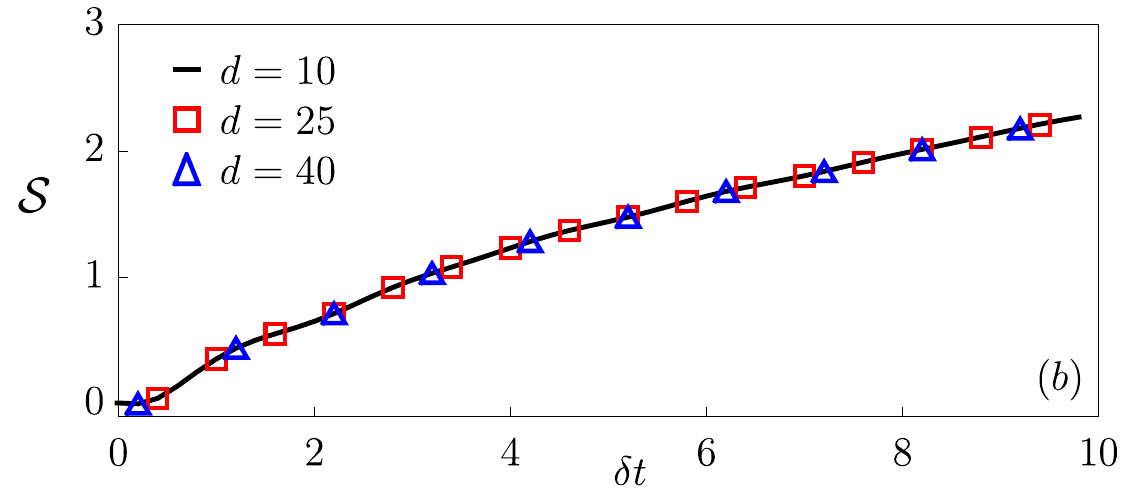}
\caption{\label{fig_entanglement_int}
\emph{
The entanglement growth for the half line $[d, \infty)$ is plotted in the non integrable spin chain ($h_z=1.1$, $h_x=0.5$) for different values of $d$ and in the subluminal case ($v=0.2$, panel $(a)$) and in the superluminal case ($v=5$, panel $(b)$).
As in the free case, a superluminal defect causes an entanglement front that never fades, but the perfect linear growth displayed in Fig. \ref{fig_EE_growth_free} (panel $(b)$) is spoiled. This can be interpreted as caused by the finite life-time of the quasiparticles induced by the intregrability-breaking perturbation.
On the horizontal axis, we pose the time lapse past after the entanglement front enters in the halfline, namely $\delta t = t -  \text{min}(d/v_\text{QP},d/v)$, elapsed after the wavefront enters the region.
}}
\end{center}
\end{figure}

We first focus on the subluminal defect $v<v_{QP}$ and consider the entanglement of two halves of the system $\mathcal{A}=[d,+\infty]$ as a function of $d$.
At the accessible times, we still observe a depletion of the propagating front (see Fig. \ref{fig_entanglement_int} $(a)$). In practice, compared with the $h_x = 0$, integrable breaking terms ($h_x\ne0$) further enhance the depletion as ballistic transport is suppressed by the finite $\tau$.
On the contrary, in the superluminal defect $v>v_{QP}$, as it happens at the integrable point, the entanglement carried by the frontwave does not exhibit any depletion increasing $d$, as it is clearly depicted in Fig. \ref{fig_entanglement_int} $(b)$. This can still be explained regarding the defect as a source of quasiparticles, however due to their finite life-time and lack of pure ballistic propagation we cannot expect linear growth any longer. However, the fact that the superluminal wavefront creates quasiparticles beyond the interval still holds true: this can be seen as a manifestation of the Cerenkov effect in this setting, which effectively enhances the entanglement growth rate.

\section{Random circuit model}
\label{sec_random}

\begin{figure}[b!]
\includegraphics[width=0.80\columnwidth]{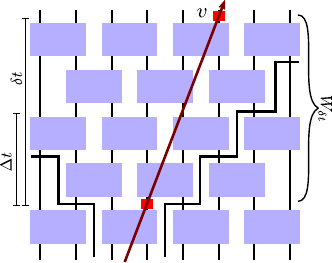}
\caption{\emph{Random circuit model: a chain of spins is evolved by applying random unitary gates in a brick-wall geometry. Each gate acting on sites $j,j+1$ is drawn from the Haar ensemble under the constraint of commuting with the local magnetization $s_j^z + s_{j+1}^z$. The action of the defect is instead encoded by using single-site random unitaries with no constraint acting on the site $j = [v t]$ every $\dt$ time steps. In this exqample $\dt = 2$ and $v = 1$.  
}
\label{fig:ruc}
}
\end{figure}

A natural question is what happens to the previous considerations for generic models where no notion of quasiparticle (not even in a perturbative sense) can be defined. Recently, \emph{random unitary circuits} (RUC) have been put forward as a new class of solvable models providing minimal and treatable examples of many-body quantum dynamics~\ocite{kpznahum,operatornahum, rucu1, rucdiffusion, RUCfloquet1, RUCfloquet2}. They are defined on a lattice of spins, in which the time evolution is performed by subsequently acting on neighbouring sites with random unitary gates drawn from an appropriate ensemble.
Here, we introduce an RUC defined of on a chain of spin $1/2$: the time evolution is performed applying gates according to the brick-wall geometry sketched in Fig.~\ref{fig:ruc}. The 2--site gates, shown as blue rectangles, represents the evolution without the defect, whose action is instead indicated with red squares. All gates are chosen independently, thus the time evolution is randomized both in space and time. However,  following~\ocite{rucu1,    rucdiffusion, rucvedika, aaron}, in order to mimic the conservation of energy away from the defect we enforce a $U(1)$ symmetry: we require that every gate acting on sites $j$ and $j+1$ commutes with $S^z_{j,j+1} \equiv s^z_j + s^z_{j+1}$. 
In practice, we denote with $\haar(n)$ the \textit{circular unitary ensemble}~\ocite{mehta} of $n\times n$ unitary matrices. The operator $\hat S^z_{j,j+1}$ has eigenvalues $S = -1,0, 1$ and 
each 2-site unitary $U$ is a $4\times 4$ matrix with a block-diagonal representation in each sector of defined $\hat S_z$, i.e. 
\begin{equation}
\label{2siteCUE}
U = \left(\begin{matrix}
     S = -1 &\\
     &    S =0 & \\
     &         &   S=1
\end{matrix}\right)
\end{equation}
with every block drawn from $\haar(d_{S})$ with $d_{\pm1} =1$ and $d_0=2$. 
Time $T$ is discrete and conventionally we take $\Delta T = 1$ for the combined action of one even and one odd layer of $2$--site unitaries (see Fig.~\ref{fig:ruc}). Because of the brick-wall geometry, 
all correlations lie inside a sharp 
lightcone with $\vmax = 2$ (black line in Fig.~\ref{fig:ruc}).

The action of the defect on site $j$ is performed via a single-site random unitary $D_{j}$. All $D_j$'s are drawn independently from $\haar(2)$, thus breaking locally the $U(1)$ symmetry. Then, the model has naturally two free parameters
\begin{itemize}
 \item $v$ is the velocity of the defect; that is, we assume that the defect acts at position $j = \round{vT}$, where $\round{x}$ indicates the nearest integer to $x$;
  \item $\dt$ is an integer representing the number of time-steps in between two different actions of the defect; in practice, the rate $\dt^{-1}$ characterizes the defect strength.
\end{itemize}
Then, the evolution of any state $\ket{\psi}$ can then be written as
\begin{equation}
\label{stateevolRUC}
 \ket{\psi(T = N \dt)} = W_{\dt}^{(N)} D_{\round{vT}}^{(N)} \dots W_{\dt}^{(2)} D_{\round{v \dt}}^{(2)} W_{\dt}^{(1)} D_0^{(1)} \ket{\psi}
\end{equation}
where $W_{\dt}$'s are defined in Fig.~\ref{fig:ruc}.

Denoting with $\overline{O}$ the average of $O$ over the ensemble of random circuits,
the computation of the average magnetization
\begin{equation}
\label{szprofdeftext}
 s^z(j,T) = \overline{\bra{\psi(T)} \hat s^z_j \ket{\psi(T)}} = \bra{\psi} \overline{\hat s^z_j(T)} \ket{\psi }
\end{equation}
can be reduced to a classical Markovian stochastic process. 
Since all unitaries are independent, in order to compute the average in Eq.~\eqref{szprofdeftext}, it is enough to analyze the average action of a single gate. If $U$ is a 2-site gate acting on the neighbouring sites $j$ and $j+1$, we have~\ocite{rucu1}
\begin{equation}
\label{twosites}
 \overline{U s_{i}^z U^\dag} =  \frac12\overline{U S^z_{j, j+1} U^\dag} = \frac12 S^z_{j, j+1}  
\end{equation}
where the first equality follows from the fact that the ensemble \eqref{2siteCUE} is invariant under the swap of sites $j,j+1$, while the second from $[U, S^z_{j, j+1}]=0$.
\begin{figure}[t!]
\includegraphics[width=1\columnwidth]{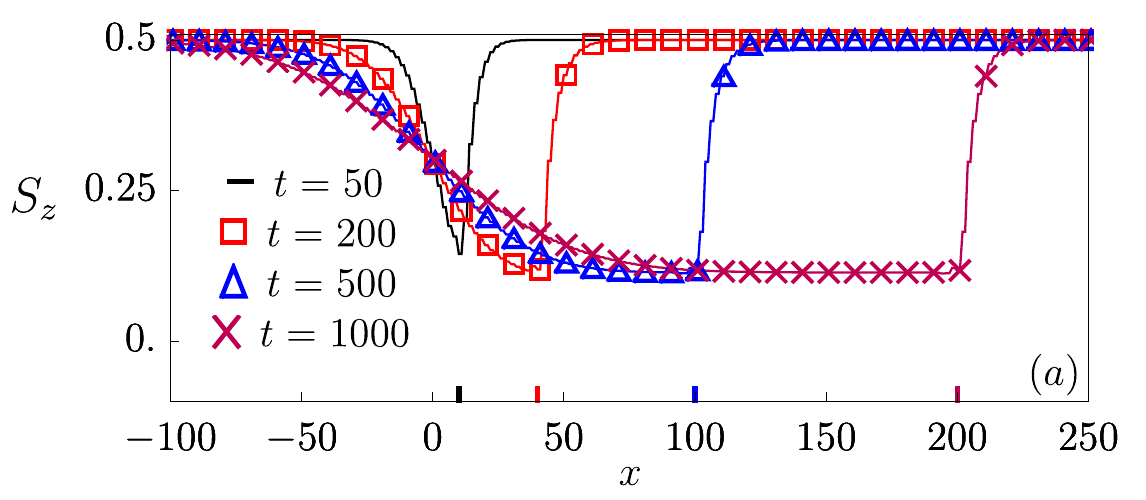}
\includegraphics[width=1\columnwidth]{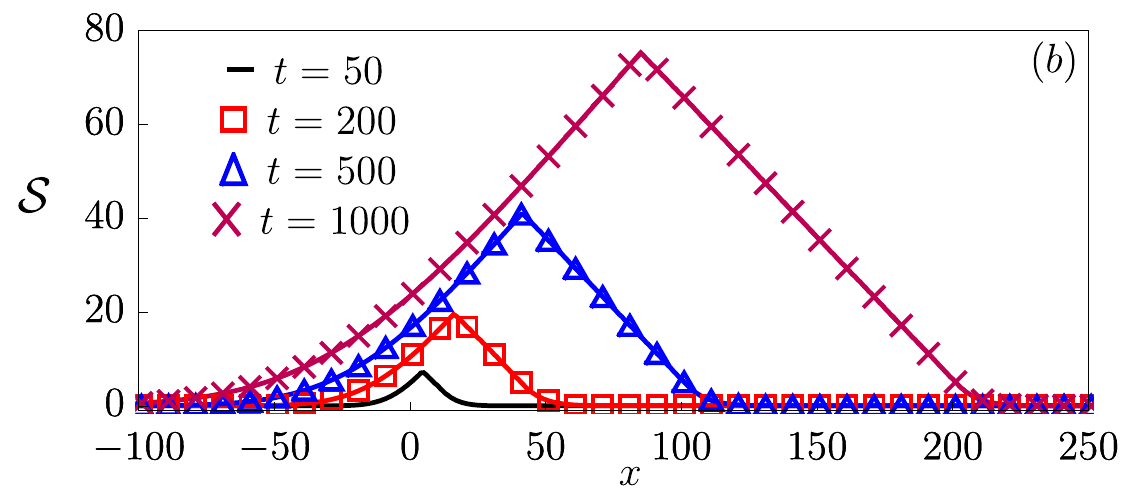}
\caption{\label{fig_RUCD}\emph{Panel $(a)$: Profile of the magnetization $s^z$ for the random circuit model with a defect moving at $v = 0.2$ and acting every $\dt = 2$ time-steps computed at different times. The dashed line is the prediction from Eq.~\eqref{analyticDiff} and $r = 3.5$ is chosen phenomenologically.
Panel $(b)$: The profile of the entanglement entropy obtained by the solution of Eq.~\eqref{subadd}}. 
On the horizontal axis, as a guide to the eye, we pose marks in correspondence with the defect's position for any $t$ we show.
}
\end{figure}
On the contrary, if $D_j$ is a 1-site random unitary corresponding to the defect action on site $j$,
\begin{equation}
\label{onesiteHaar}
 \overline{D_j s_{i}^z D^\dag_j} =  (1 - \delta_{i,j}) s_{i}^z\, .
\end{equation}
Eqs.~\eqref{twosites} and \eqref{onesiteHaar} completely characterize the Heisenberg evolution of local magnetizations once averaged over the random circuit ensemble. 
In particular, they imply a linear relation
\begin{equation}
\label{linearevol}
 \overline{\hat s_j^z(T = N \dt)} = \sum_{j'} \mathcal{M}_{j,j'}(T)  \hat s_{j'}^z
\end{equation}
where the matrix $\mathcal{M}(T)$ can be decomposed as a matrix product
\begin{equation}
 \mathcal{M}(T) =  \mathcal{W} \; \mathcal{D}_{\round{v  T}} \dots  \mathcal{W} \; \mathcal{D}_{\round{v \dt}} \mathcal{W} \; \mathcal{D}_0
\end{equation}
where we defined $ (\mathcal{D}_i)_{j,j'} = \delta_{j,j'} (1 - \delta_{i,j})$, while the explicit form of $\mathcal{W}$ can be obtained from the repeated action of \eqref{twosites} over a sequence of $\dt$ even and odd layers (see Fig.~\ref{fig:ruc}). 
After the quantum average of \eqref{linearevol} over the initial state $\ket{\psi}$, 
we get an exact expression for the magnetization profile
\begin{equation}
\label{linearrelsz}
 s^z(j,T) = \sum_{j'} \mathcal{W}_{j,j'}(T) s^z(j', 0) \;.
\end{equation}
We are interested in an initially weakly entangled state, so for simplicity we focus on the completely polarized state along $z-$direction, i.e. $\ket{\psi} = \ket{\ldots \uparrow \uparrow \ldots}$, which reproduces some features of the groundstate considered before and $s^z(j, 0) = 1/2$.
Indeed, thanks to the $U(1)$ symmetry, this is an invariant state under the time evolution without the defect. In practice, the defect behaves as a moving source of magnetization. 

In order to gain some insights about the dynamics induced by Eq.~\eqref{linearrelsz}, one can perform a long-wavelength expansion to get a coarse-grained continuous description. More simply, as explicitly shown in [\onlinecite{rucu1}], we observe that Eq.~\eqref{twosites} describes an unbiased random walk which is clearly described by the diffusion equation in the continuous limit. Then, according to Eq.~\eqref{onesiteHaar}, the defect acts by locally removing the present magnetization. In the continuous limit $j \to x, T \to t$, this suggests the form
\begin{equation}
\label{diffusiondef}
 \partial_t s^z(x, t) = D \partial_{xx} s^z(x,t) - \frac{v r}{D} \delta(x - v t) s^z(x,t) \;.
\end{equation}
The diffusion constant can be determined directly from \eqref{twosites} and one finds $D = 1$ in our unities. The second term in the right-hand side of \eqref{diffusiondef} accounts for the magnetization removed by the defect action. The dimensionless parameter $r$ controls the defect strength and depends in a non-trivial way on $v$ and $\delta t$, because the behavior of $s^z(j,t)$ jumps erratically around $j \sim v t$ due to lattice effects. Its precise value is unnecessary for our analysis, though we note that for $v > \vmax$, one has simply $r = (v \dt - 1)^{-1}$, which is obtained matching the global magnetization change between the lattice and continuous descriptions. 
Setting $s^z(x,t) = \frac12 - n(x,t)$, with $n(x,t)$ the local density of spin flips, Eq.~\eqref{diffusiondef} with $s^z(x,t=0) = 1/2$ is solved at large times by
\begin{equation}
\label{analyticDiff}
n(x,t) = \begin{cases}
             \frac{r \exp[-v (x-vt)/D]}{2(r+1)} & x \gtrsim v t \\
             \frac{r(1 + \operatorname{Erf}\left(x/\sqrt{4 D  t}\right))}{4 (r+1)} & x \ll vt
            \end{cases} \;,
\end{equation}
i.e. it is characterised by a broadening front at the initial defect position plus a travelling wave dragged by the defect. Interestingly, the defect is preempted by an exponentially decaying front on the finite length scale $\sim D/v$.  
As shown in Fig.~\ref{fig_RUCD}~left, this coarse-grained description  captures well the magnetization profile for large times and small $v$'s. However, in this continuous limit, all lattice effects are washed out and no role is played by $\vmax$.

A more refined description of the tails of the magnetization profile can be obtained via a large deviation Ansatz, i.e. $s^z(j,T) \simeq 1/2 - e^{- T \phi(j/T)}$. In other words, on each fixed ray $j/T = \zeta$, we define
 \begin{equation}
 \phi(\zeta) \equiv - \lim_{T\to\infty} \frac1T\log(n(\zeta T, T))  \;.
\end{equation}
An expression for $\phi(\zeta)$ can be derived from the explicit solution of Eq. ~\eqref{szprofdeftext}, as we now discuss.
First, by acting explicitly with $\mathcal{D}_j$ in \eqref{linearevol} and neglecting lattice discretization, we can rewrite
\eqref{linearevol} as
\begin{equation}
\label{sumn}
n(j,T) = \mathcal{W}_{j, v T} (1/2 - n(v T, T-\dt))+ \sum_{j'} \mathcal{W}_{j,j'} n(j', T-\dt)\, .
\end{equation}
For any $v \neq 0$ and large $T$, the term $n(v T, T -\dt) \sim e^{- T \phi(v)}$ is exponentially small and thus negligible with respect to the factor $1/2$. The 
resulting equation can then be solved by iteration, leading to
\begin{equation}
 n(j,T) \sim \frac 12  \sum_{k=1}^{T/\dt} [\mathcal{W}]^k_{j, v(T - k\dt)}\, .
\end{equation}
Setting $k = T u/\dt$, we can turn the sum into an integral by using that
the $k$-th power of the matrix $\mathcal{W}$ can be written explicitly as \ocite{rucu1}
\begin{equation}
 [\mathcal{W}]^{k}_{j,j'} \stackrel{k = Tu/\dt}{=} 2^{-2 u T} \binom{2 u T-1}{u T+\left\lfloor \frac{j-j'}{2}\right\rfloor } \simeq \exp\left[- T u I\Bigl(\frac{j'-j}{u T}\Bigr)\right]
\end{equation}
where for $|z|\leq 2$
\begin{equation}
 I(z) = 
\left(\frac{z}{2}+1\right) \log \left(\frac{z}{2}+1\right)+\left(1-\frac{z}{2}\right) \log \left(1-\frac{z}{2}\right) \;, 
\end{equation}
and $I(z) = \infty$ otherwise. 
We thus get
\begin{equation}
 n(z T,T) \sim \int_0^1 du \; \exp\left[-T u I\Bigl(\frac{v (1 - u) - z}{u}\Bigr) \right]\, .
\end{equation}
At large $T$, the integral can be evaluated by saddle point and leads to
\begin{multline}
\label{ldsol}
 \phi(z) = \min_{u \in [0,1]} \left[ u I\Bigl(\frac{v (1 - u) - z}{u}\Bigr)\right] =\\ 
 =\begin{cases}
  I(z) \;, & z < 0 \cap z > z_c \\
  0 \;, & 0<z<v \\
  (z-v)\frac{I(z_c)}{z_c - v}  \;, & v <z < z_c
 \end{cases}
 \end{multline}
For $v < \vmax = 2$, the value $z_c$ is determined by the equation
\begin{equation}
 (z_c - v) I'(z_c) = I(z_c)  \;, \quad z \in [v, \vmax]
\end{equation}
while $z_c = v$ for $v > \vmax$. 

\begin{figure}[t!]
\includegraphics[width=1\columnwidth]{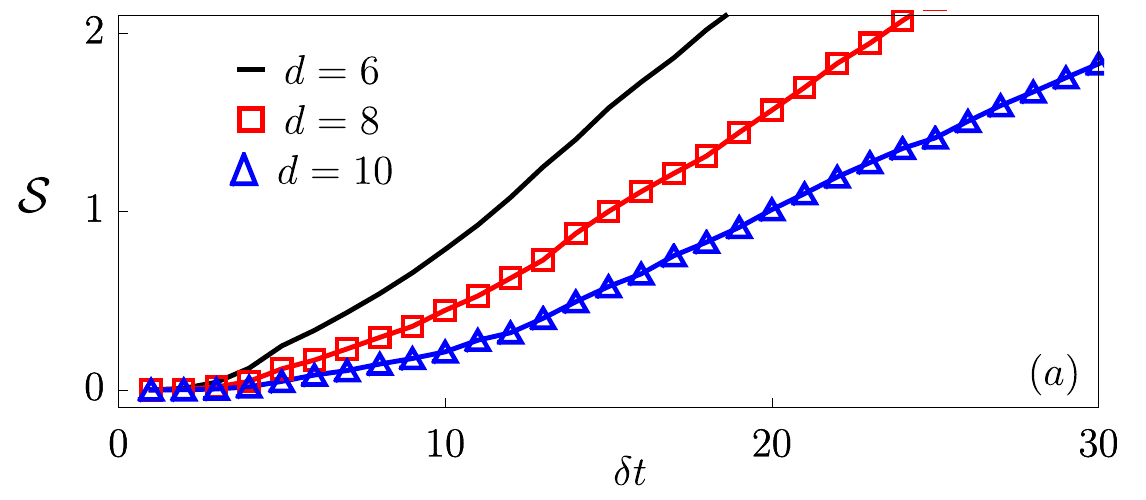}
\includegraphics[width=1\columnwidth]{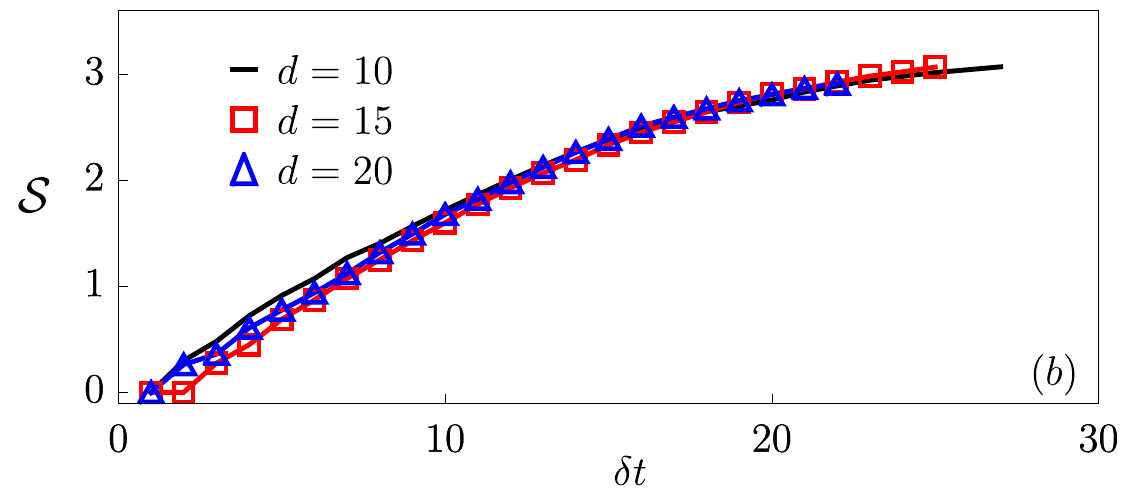}
\caption{\emph{Time-dependence of entanglement entropy for the interval $[d,\infty)$ for $v = 0.2 < \vmax$ $(a)$ and $v = 3.0 > \vmax$ $(b)$. On the horizontal axis, the time delay $\delta t = t - d/\max(\vmax, v)$ from when the front hits site $d$ is considered. 
Once again, the entanglement front fades in the subluminal case, while it persists unscathed for the superluminal defect.
Data are obtained by MPS simulations averaged over 100 samples. }\label{entRUC}}
\end{figure}

\begin{figure*}[ht!]
\begin{minipage}[l]{\textwidth}
\begin{center}
\includegraphics[width=1\textwidth,valign=l]{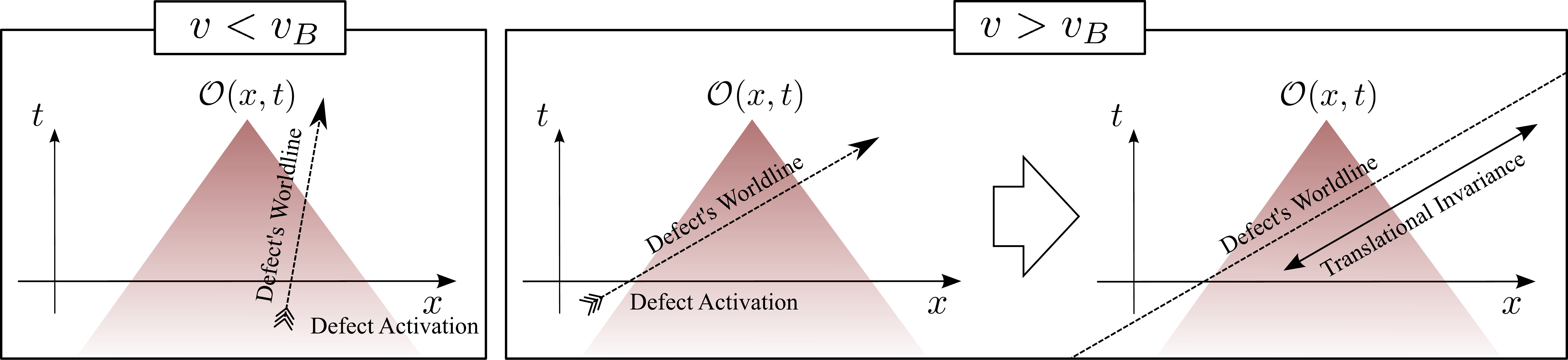}
\caption{\label{fig_lightcone}
\emph{The sudden formation of the comoving steady state. In the subluminal case (left), the causal lightcone of a local observable $\mathcal{O}(x,t)$ in the space-time plane is such that, if it is crossed by the defect worldline, then it also contains the defect-creation event. Conversely, in the superluminal case, local observables right behind the defect are such that their lightcone is crossed by the defect worldline, but the defect-creation event lays outside of the latter. This is therefore equivalent to a setting where the defect was located infinitely back in the past. Therefore, translational symmetry along the defect's worldline $\langle \mathcal{O}(x,t)\rangle=\langle \mathcal{O}(x+v\Delta t,t+\Delta t)\rangle$  emerges.	
}
}
\end{center}
\end{minipage}
\end{figure*}

In practice, as it happens in the presence of quasiparticles, for subluminal defects ($v<\vmax$), a smooth front for $x \in [vt, \vmax t]$ foreruns the defect. However, contrarily to the integrable case, here the decay of the evanescent front is exponentially fast being due to a large deviation of an otherwise diffusive dynamics. This quick equilibration is a hallmark of chaotic dynamics. 
On the contrary, for $v > \vmax$, the magnetization profile jumps abruptly around the $x \sim vt$ as the system has no time to equilibrate the excess of magnetization produced by the defect. 

These considerations reflect in the front of entanglement carried by the defect. Unfortunately, the exact calculation of the entanglement entropy is a non-trivial task for RUC~\ocite{rucdiffusion, entgrowth2019}. Nonetheless, an exact upper bound is provided by subadditivity~\ocite{nielsenchuang, kpznahum,entgrowth2019}. In particular, at any time it must hold $\svn(j + 1) \leq \svn(j) + \mathfrak{s}(j)$, being $\svn(j)$ the entanglement entropy for the half-line $[j, \infty)$ and $\mathfrak{s}(j)$ the one of the single spin at site $j$ with the rest of the system. 
Following \ocite{entgrowth2019}, we assume that local equlibration has occurred on the scale of single sites. Therefore, $\mathfrak{s}(x,t) \sim \mathfrak{s}(x+1,t) \sim \eta(n) \equiv - n \ln n - (1-n)\ln(1-n)$. Using subadditivity from the two sides, we have the update rule whenever a 2-site unitary is applied on the bond $x$
\begin{align}
\label{subadd}
 \svn(x,t+1) \lesssim \min[\svn(x-1, t), \svn(x+1,t)] + \eta(n)\, .
\end{align}
Note that the action of the defect cannot directly change $\svn(x,t)$, but it affects the profile of $n(x,t)$ which enters in \eqref{subadd}. Interpreting this inequality as an equality, it gives an update rule for the entropy at any position, which depends on the local density $n(x,t)$. 
This approach was originally used in \ocite{kpznahum} to show the emergence of the Kardar-Parisi-Zhang equation in the entanglement dynamics without conserved quantities. Then, in its form \eqref{subadd}, it was recently applied in \ocite{entgrowth2019} for several inhomogeneous setups. Here, we apply it to the moving defect model. Although it only provides an upper bound, we expect it to capture the qualitative behavior of the entanglement dynamics. The result for the subluminal case is shown in Fig.~\ref{fig_RUCD}~(b).

We can use Eq.~\eqref{subadd} to get a qualitative estimate of the entanglement front. 
We fix a large $d$ and look at the time dependence of $\svn(d, t = x/v_{F} + \delta t)$ with $v_F = \max[\vmax = 2, v]$ the front velocity.
Assuming $\svn(x-1,t) \sim \svn(x+1,t)$, we have $\partial_t s(x,t) \sim \eta(n) \stackrel{n\ll1}{\sim} n$. Using \eqref{ldsol}, we get for $v < \vmax$, $\svn(d, d/\vmax + \delta t) \sim 2^{-d} d^{\delta t}$. On the contrary, for a superluminal defect $s(d, d/v + \delta t) \propto \delta t$, independently of $d$. 
These qualitative predictions are verified with MPS simulations performed on the RUC in Fig.~\ref{entRUC}. 

In conclusion, the chaotic model described by the RUC has a phenomenology similar to the integrable case, although the decay of the endogenous entanglement front is much faster and signals the onset of diffusive behavior of transport. Note that this is not in contrast with the ballistic propagation of information observed in chaotic diffusive systems~\ocite{ballistic2013}: here we start from the fully polarized state and therefore, in the absence of magnetization transport, there is no local entropy available for entanglement to grow.

\section{The comoving steady state}
\label{sec_comovingtail}

In Sec. \ref{sec_quasiparticles} we saw how the exogenous entanglement front due to superluminal defects in integrable models can be framed within the quasiparticle picture, which holds true even in presence of weakly integrable breaking perturbations, despite the quasiparticles acquire a finite lifetime.
In Sec. \ref{sec_random} we considered random circuits, where no quasiparticle interpretation is known, finding again the same features in the entanglement front of superluminal defects. In this respect, one could suspect a very general argument should exists, based solely on the existence of a maximum velocity.
Indeed, this this is the case, as we further argument below.
Beyond the presence of a persistent entanglement front,
the existence of $\vmax$ is associated with stationarity in the reference frame co-moving together with the defect. In particular, one can define the butterfly velocity $\vB < \vmax$, associated to the spreading of a local operator under Heisenberg evolution: the support of $\mathcal{O}(x,t)$ is contained in $[x - \vB t, x + \vB t]$ up to exponentially small error~\ocite{diffusivemetals2017,cit_lightcone, operatornahum}. 
Note that although the butterfly effect is normally associated with chaotic systems, 
operators exhibit spreading also for integrable models, with $\vB \sim \vqp$~\ocite{nahumlyapunov}. 

The formation of a comoving steady state is best appreciated through a simple relativistic argument, for which we refer to Fig. \ref{fig_lightcone}.
Borrowing a relativistic terminology, we dub as time-like the inside of the lightcone spreading with velocity $\vB$ from the initial position of the defect, while points placed outside of it will be called space-like.
Because information propagates at a finite velocity, any measurement performed at a space-like point $(x,t)$ can only be affected by the causal lightcone which ends at this point, as shown in Fig.~\ref{fig_lightcone}. If the perturbation is subluminal, whenever its wordline crosses the causal lightcone, also the space-time point $(x=0, t=0)$ associated with the formation of the perturbation is contained in the causal lightcone. On the contrary,
if the perturbation is travelling at $v>v_B$, it can enter the causal lightcone, even though
the instant of its creation remains always outside (see Fig.~\ref{fig_lightcone} -- right).
\begin{figure}[b!]
\begin{center}
\includegraphics[width=0.9\columnwidth,valign=l]{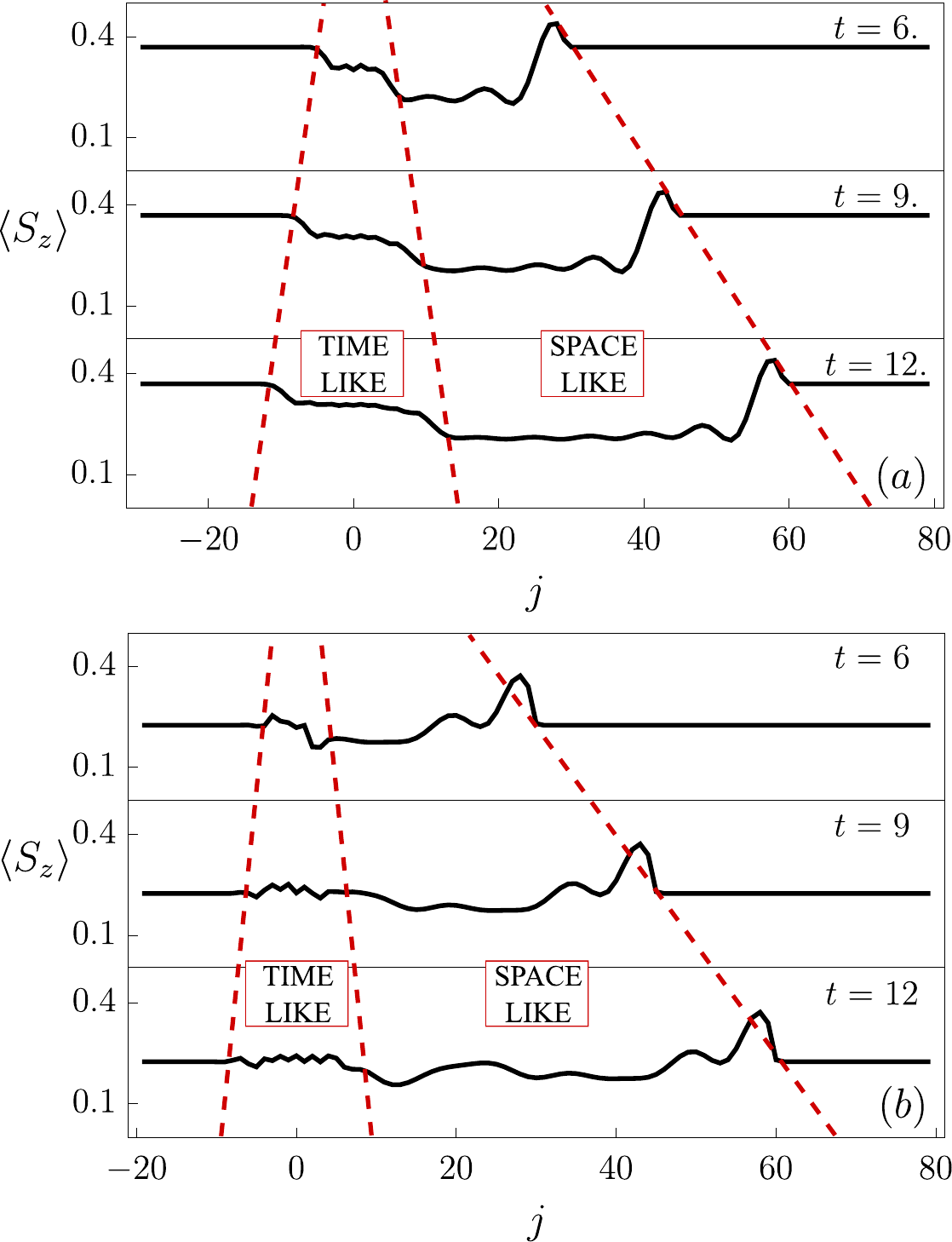}
\caption{\label{fig_tail}\emph{Magnetization tail formation behind a superluminal defect ($v=5$, $h_z=1.1$ ) at the integrable (Panel $(a)$ $h_x=0$) and non integrable point (Panel $(b)$ $h_x=0.5$). Within the time-like lightcone, the dynamics is rather different in the two cases, due to the presence of ballistic transport in the first case and lack of it in the second one. Instead, within a space-like region ranging from the time-like lightcone until the defect position, in both cases a non trivial tail steadily following the defect is formed.}}
\end{center}
\end{figure}
This implies that, while a local measurement can be affected by the perturbation, its creation in the space-time plane remains unknown to any local observer and can be equivalently thought to be located infinitely back in the past. In this case, the translational symmetry along the defect worldline emerges and the expectation value of the local observable remains unchanged moving parallel to it. In other words, the expectation value is stationary in the comoving reference frame. 
In Fig. \ref{fig_tail} we probe the described general framework, testing the profile of the local magnetization at different times after the defect activation, both in the integrable and weakly non-integrable case, finding perfect agreement with the described scenario. For the RUC model, due to the sharp lightcone, stationarity in the comoving frame is granted after ensemble average has been considered. 

The existence of this stationary state also explains the behavior of the entanglement front. Indeed, for low entangled states, namely possessing a finite correlation length, the entanglement content is determined by the local properties of the state. The latter is fully determined by the position of the defect with respect to the interval of interest and not on the actual time. This is of course true as long as the interval $[d,d+\ell]$ lays outside of the causal lightcone spreading at velocity $v_\text{B}$ from the space-time point where the defect was activated.

\newpage

\section{Conclusions}
\label{sec_conclusions}

We theoretically investigate and numerically confirmed universal features of isolated quantum systems probed with moving impurities. The presence of a maximum velocity in the information spreading dramatically affects the system response, with direct experimental implications.
Firstly, the evolution of entanglement is nowadays measurable in cold-atomic experiment \ocite{greiner2015}.
Secondly, the formation of a stationary trail which follows the defect is independent on any fine tuning of the model, relying only on the existence of a finite $v_B$, making it an ideal candidate to be observed in actual experiments. 
Finally, we expect a similar phenomenology to emerge in higher dimensions, with the additional intriguing possibility to observe a \v Cerenkov angle in the entanglement propagation front. Such experimental ways to investigate quantum chaos could be insightful to understand this fascinating problem.

\acknowledgments
We are grateful to Mario Collura, Pasquale Calabrese, Fabian Essler for useful discussions.
A.B. acknowledges the support from the European
Research Council under ERC Advanced grant
743032 DYNAMINT.
The work was supported in part by the European Union's
Horizon 2020 research and innovation programme under
the Marie Sklodowska-Curie Grant Agreement No.
794750 (A.D.L).
The matrix-product state simulations were performed using the open-source ITensor library.

\appendix

\section{ Numerical methods}
\label{app_numerics}
\subsection{Green function transfer matrix}

As long as the free point in the Ising chain is considered $(h_x=0)$, the mapping to the fermionic basis can be exploited in the numerical solution and considerable large times and system's sizes reached ($\sim 1500$ lattice sites are an easy task for standard laptops).
The same algorithm has also been used in Ref. \ocite{AA2018ising}, but nevertheless we discuss it hereafter for the sake of completeness.

It is useful to reformulate the linear Heisenberg equation of motion within a Green function approach. In this perspective, the fermionic field $\psi$ \eqref{eq:mode} at time $t$ is connected to the field at time $t_0$ by mean of a Green function

\be\label{greenfermion}
\psi_j(t)=\sum_{j'} G_{j,j'}(t,t_0)\psi_j(t_0)\, ,
\ee
The Green function $G_{j,j'}(t,t_0)$ for each value of the indexes $(j,j')$ is a $2\times 2$ matrix that is required to solve the linear differential equation
\begin{multline}
i\partial_tG_{j,j'}(t,t_0)=\frac{(i\sigma^y-\sigma^z)}{2}G_{j-1,j'}(t,t_0)+\\
-\frac{(i\sigma^y+\sigma^z)}{2}G_{j+1,j'}(t,t_0)+\sigma^z [h_z+V(j-vt) ]G_{j,j'}(t,t_0)\label{dG}\, ,
\end{multline}
where $\sigma^{x,y,z}$ are the standard Pauli matrix. For $t=t_0$, the Green function must be required to be the identity
\be
G_{j,j'}(t_0,t_0)=\delta_{j,j'}\begin{pmatrix} 1 && 0 \\ 0 && 1 \end{pmatrix}\, .
\ee
Obviously, we are ultimately interested in the case $t_0=0$: the knowledge of the Green function links the local observables at time $t$ with those at the initial time (i.e. the analytically known correlators on the Ising ground state).
In particular, let us introduce a two dimensional vector
\be
\psi_j=\begin{pmatrix} d_j \\ d^\dagger_j \end{pmatrix}
\ee
and  therefore the correlation matrix of the fermions can be written as $\langle \psi_j\psi^\dagger_{j'}\rangle$
\be\label{two_point}
\langle \psi_j\psi^\dagger_{j'}\rangle=\begin{pmatrix} \langle d_jd^\dagger_{j'}\rangle && \langle d_jd_{j'}\rangle \\ \langle d^\dagger_{j}d^\dagger_j\rangle && \langle d^\dagger_jd_{j'}\rangle\end{pmatrix}\, .
\ee
Then Eq. \eqref{greenfermion} simply implies
\be\label{evocorr}
\langle \psi_j(t)\psi^\dagger_{j'}(t)\rangle=\sum_{l,l'}G_{j,l}(t,0)\,\,\langle \psi_l(0)\psi^\dagger_{l'}(0)\rangle\, \,G_{j',l'}^\dagger(t,0)\, .
\ee

Rather than solving directly the differential equation for the Green function, we can conveniently take advantage of its composition properties. Indeed, the Green function must obviously satisfy
\be
G_{j,j'}(t_3,t_1)=\sum_{l}G_{j,l}(t_3,t_2)G_{l,j'}(t_2,t_1)\, .\label{compG}
\ee

Because of the moving defect, time translational invariance is broken and $G_{j,j'}(t,t_0)$ has a non trivial dependence on both times (and not only on the difference $t-t_0$ as it would happen in the absence of the defect). However, the fact that the defect moves at constant velocity provides a periodicity in the Green function.

In fact, after a time $1/v$, the defect shifts of one site: time translations of steps $1/v$ can be equivalently regarded as translations on the lattice
\be
G_{j,j'}(t_2+nv^{-1},t_1+nv^{-1})=G_{j-n,j'-n}(t_2,t_1)\, , \label{shG}
\ee
for any integer $n$.
By mean of a combination of Eq. (\ref{compG}) and Eq. (\ref{shG}), we can readily write a recurrence relation obeyed by the Green function
\begin{multline}
G_{j,j'}(2^n v^{-1},0)=\sum_l \Big(G_{j-2^{n-1},l-2^{n-1}}(2^{n-1} v^{-1},0)\times \\G_{l,j'}(2^{n-1} v^{-1},0)\Big)\, .\label{recG}
\end{multline}
While each step of the recurrence relation requires a matrix product (computationally expensive), large times can be reached exponentially fast, provided the first step $G_{j,j'}(v^{-1},0)$ is known.

Concerning the computation of  $G_{j,j'}(v^{-1},0)$, we can reason as it follows: at time $t=0^+$ the defect is assumed to be right beyond the lattice $j=0$. Until a time $t=1/v$ is reached, the system evolves freely as if it was homogeneous: the homogeneous Green function $G^0_{j,j'}$ is easily exactly computed in terms of the modes of the free Ising chain
\begin{multline}
G^0_{j,j'}(t,0)=\frac{1}{N}\sum_{m}^N e^{i2\pi m (j-j')/N}\times\\
\Big[e^{-i\omega(2\pi m/N) t} u_1(2\pi m/N)u_1^\dagger(2\pi m/N)+ \\
+e^{i\omega(2\pi m/N) t} u_2(2\pi m/N)u_2^\dagger(2\pi m/N)\Big]\, ,
\end{multline}
where the sum is over the (half)integers up to the total number of lattice sites $N$, depending on being in the (even)odd magnetization sector. The vectors $u_{1,2}$ are defined in Eq. \eqref{eq_udef}, while $\omega$ is the Ising dispersion law.

At time $t=1/v$ the defect suddenly kicks the system and the Green function has a jump dictated by the singular term in the Schr\"oedinger equation \eqref{dG}
\be
G_{j,j'}(v^{-1},0)=e^{-i\frac{c}{v}\delta_{j,1}\sigma^z}G^0_{j,j'}(v^{-1},0)\, .
\ee
This concludes the computation of $G_{j,j'}(v^{-1},0)$, which can now be employed in the recurrence relation \eqref{recG}.
Once the Green function has been computed, the correlation functions easily follows through \eqref{evocorr}: finally, from the correlators the Entanglement Entropy of intervals can be obtained taking advantage of the gaussianity of the model (see eg. Ref. \ocite{fagottiXY}). The algorithm here presented makes possible to sample times $t=2^n/v$ with machine precision, but suitable generalizations allow for a ticker time sampling at the price of introducing more matrix products.

\subsection{Matrix-product states simulation}
\subsubsection*{Ising model}
Numerical simulations for the Ising Hamiltonian Eq. \eqref{Hising} in the presence of the integrability-breaking term $h_x \neq 0$ were performed in two steps:
\begin{enumerate}
\item a representation of the initial state $\ket{\Psi}$ (the groundstate of the Hamiltonian for $\kappa = 0$) was obtained using the DMRG algorithm using $10$ sweeps as a matrix-product state (MPS)  with maximal bond dimension $\chi = 200$;
\item the time evolution for each time step $\Delta t = 1/v$ was performed alternating evolution without the defect and the action of the defect on a single site. 
This leads after $n$ time steps to
\begin{equation}
\label{discreteIsing}
\ket{\Psi(t = n \Delta t)} = \hat D_n \hat  W_{\Delta t} \ldots D_2 \hat W_{\Delta t} \hat D_1 \hat W_{\Delta t} \ket{\Psi(t=0)} 
\end{equation}
where $\hat W_{\Delta t} = e^{-\imath \hat H_{\kappa=0} \Delta t}$ is the time-evolution in the absence of the defect (when the defect lies in between two lattice sites) and $D_j = e^{- \imath \kappa  \sigma_j^z/v}$ is the action of the defect on the site $j$. In order to implement the two steps:
\begin{enumerate}
 \item we used a matrix-product operator (MPO) approximation $\tilde W_{\Delta t} \simeq W_{\Delta t}$ using the method described in \ocite{karrasch2015}. The unitary operator $\tilde W$ was then applied to the state $\ket{\Psi(t)}$ and the result was recompressed as a new MPS by discarding all Schmidt eigenvalues smaller than $\delta\lambda$; to increase the precision, the time interval $\Delta t$ was splitted into $N$ smaller steps, i.e. $W_{\Delta t} = W_{\Delta t/N}^N$, so that $\Delta t/N < 10^{-3}$. The action of each $W_{\Delta t/N}$ was approximated with the procedure above. Moreover, two complex time steps were used to further reduce the scaling of the errors with $\Delta t/N$ (see \ocite{karrasch2015} for details). 
 \item the action $\hat D_j$ of the $\delta$-defect was implemented acting on the local Hilbert space of the site $j$. 
\end{enumerate}
\end{enumerate}
The validity of the method was benchmarked by comparison with non-interacting case. In general, the accuracy was kept under control by considering two different truncation errors $\delta \lambda = 10^{-9}$ and $\delta \lambda = 10^{-12}$. The two values of $\delta \lambda$ always provided comparable results (the difference being smaller than the symbols in the plot of Fig.~\ref{fig_tail}) for all the times in the simulation. Two different simulations were run 
with maximum limit for the bond dimension set to $\chi_{max} = 250$ and $\chi_{max} = 500$. The simulation was stopped when the two simulation showed significative disagreement.

\subsubsection*{Random unitary circuit}
The numerical treatment of the RUC model described in Sec.~\ref{sec_random} is formally analogous to the treatment of the Ising model described above. Indeed Eq.~\eqref{discreteIsing} shows manifest analogies with Eq.~\eqref{stateevolRUC}. The main difference is that, thanks to the brick-wall structure in Fig.~\ref{fig:ruc}, the factors $W_\tau^{(N)}$ are automatically expressed as products of local $2$--site gates, which can be easily applied to an MPS. In order to sample from the CUE, we used the algorithm in [\onlinecite{mezzadri}]. 

%%%%%%%%%%%%%%%%%%%%%%%%%%%%%%%%%%%%%%%%%%%%%%%%%%

\end{document}